\documentclass[aps,twocolumn,showpacs,preprintnumbers,amsmath,amssymb,prl]{revtex4-1}
\usepackage{epsf,amsmath,amssymb,verbatim,color,multirow,pifont}
\usepackage{graphicx}

\begin{document}
\title{Discontinuous percolation transitions in real physical systems}
\author{Y.S. Cho and B. Kahng}
\affiliation{{Department of Physics and Astronomy, Seoul National University,
Seoul 151-747, Korea}}
\date{\today}

\begin{abstract}
We study discontinuous percolation transitions (PT) in the diffusion-limited cluster aggregation model of the sol-gel transition 
as an example of real physical systems, in which the number of aggregation events is regarded as the number of bonds occupied in the system. When particles are Brownian, in which cluster velocity depends on cluster size as $v_s \sim s^{\eta}$ with $\eta=-0.5$, a larger cluster has less probability to collide with other clusters because of its smaller mobility. Thus, the cluster is effectively more suppressed in growth of its size. Then the giant cluster size increases drastically by merging those suppressed clusters near the percolation threshold, exhibiting a discontinuous PT. We also study the tricritical  behavior by controlling the parameter $\eta$, and the tricritical point is determined by introducing an asymmetric Smoluchowski equation. 
\end{abstract}

 \maketitle

Percolation, a stochastic model for the formation of macroscopic-scale spanning clusters, has received considerable attention in statistical physics 
for a long time as a model for metal-insulator transitions, sol-gel transitions, epidemic spreading, fracture and so on~\cite{stauffer}.
When a control variable, which is the occupation probability of a conducting bond between two vertices, is increased, a long-range spanning
cluster emerges at a critical threshold $p_c$. Such a percolation transition (PT) is conventionally continuous. The discovery of a {\it discontinuous} PT has therefore been a long-standing issue in statistical physics. In this circumstances, a recently introduced stochastic model~\cite{science} for the explosive PT has attracted considerable attention in a short time period. This stochastic model is a simple modification of the classical Erd\H{o}s-R\'enyi (ER) random graph model~\cite{er}, which contains suppression effect in growth of cluster sizes. Such an explosive PT behavior has also been observed in other recently introduced stochastic toy models~\cite{friedman, fss, can, hamiltonian, manna, souza, herrmann}. Even though 
it has been controversial if such an explosive PT is indeed discontinuous in the thermodynamic limit ~\cite{mendes,hklee,science2}, the explosive percolation model has opened a new avenue for the study of discontinuous PT in non-equilibrium systems. In this Letter, we are interested in what a physical system in real world a discontinuous PT can be observed. 

In this Letter, we consider the diffusion-limited cluster aggregation (DLCA) model~\cite{dla, meakin, kolb} for the sol-gel transition as a candidate in real-world systems of showing a discontinuous PT. This model was introduced a long time ago, and the dynamic cluster-size distribution were intensively studied for this model~\cite{vicsek, fdim, cdist, flory, experiment}. Here, this model is studied in the context of PT, which takes place as the number of cluster aggregation events increases. We also show that indeed the DLCA model exhibits a  discontinuous PT. Furthermore, we generalized the DLCA model in which cluster velocity depends on cluster size as $v_s \sim s^{\eta}$. As the parameter $\eta$ varies, there exists a tricritical point beyond which the PT becomes continuous with continuously varying exponents. We show that the generalized DLCA model can be represented via an asymmetric Smoluchowski equation, by which the tricritical point can be determined.

The DLCA model is simulated in the following way: Initially, $N$ single particles are placed randomly in $L\times L$ square lattices. Simulations
start from $N$ mono particles. The density of the particles is given as $\rho=N/L^2$. The system size $L$ is controllable, while the density remains fixed in the simulations. Here we consider the case that particles are Brownian, so that velocity of a cluster is inversely proportional to the square root of its size~\cite{reif}. To implement, we perform simulations as follows~\cite{kolb}: At each time step, (i) a $s$-sized cluster is selected with the probability $q\equiv s^{-0.5}/(\sum_s N_s s^{-0.5})$, and is moved to the nearest neighbor. When two distinct clusters are placed at the nearest-neighbor positions, these clusters are regarded as being merged, forming a larger cluster. (ii) The time is advanced by $\delta t=1/(\sum_s N_s s^{-0.5})$, where $N_s$ is the number of $s$-sized clusters. We iterate the steps (i) and (ii) until the giant cluster size is $N$. Later, we will consider a more general case in which the velocity is proportional to $s^{\eta}$~\cite{cdist}. 

\begin{figure}[t]
\includegraphics[width=0.8\linewidth]{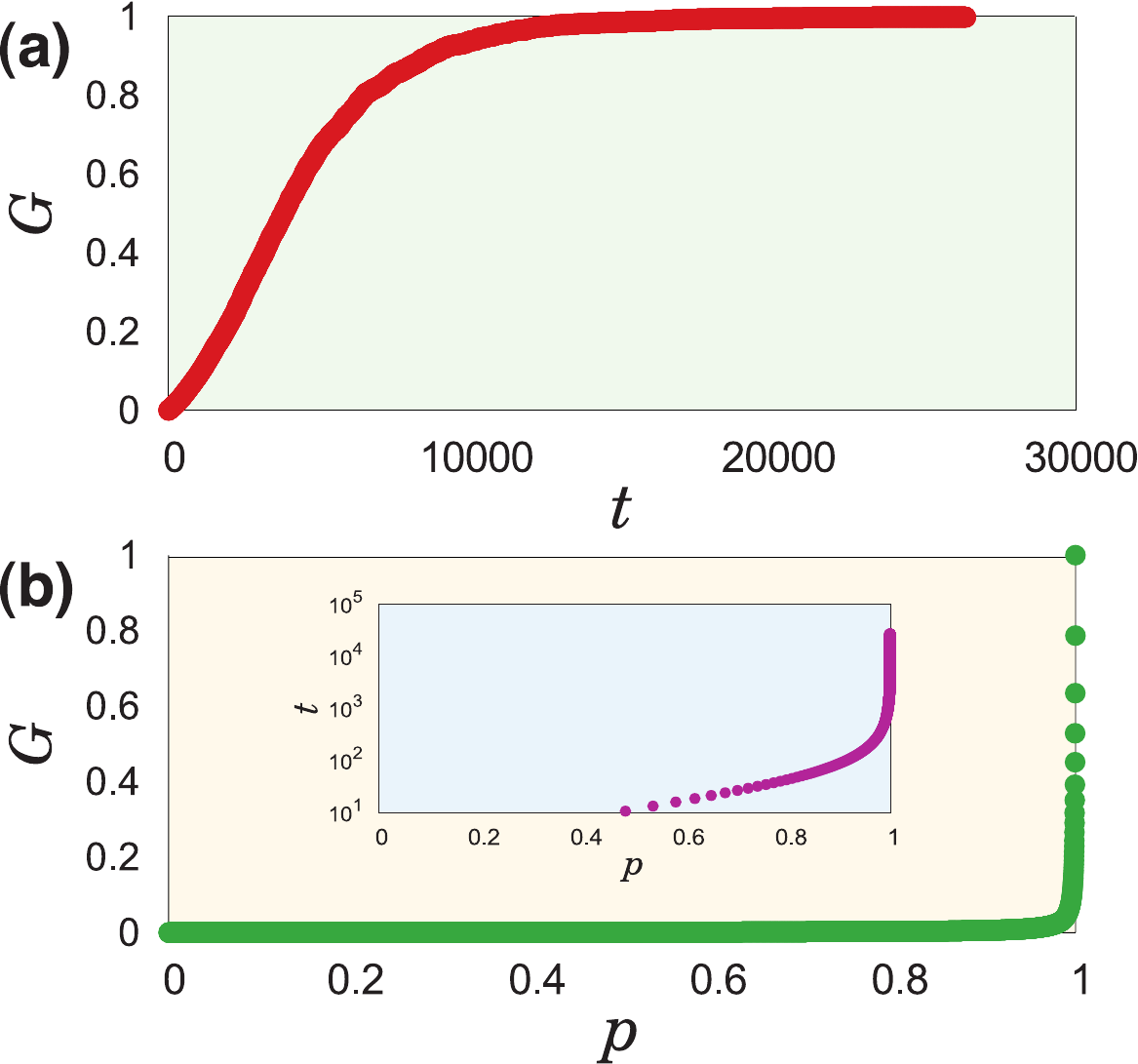}
\caption{(color online) (a) Plot of the giant cluster size $G$ versus $t$ when particles are Brownian. The giant cluster grows continuously from
$t=0$. (b) Plot of $G$ versus $p$. $G$ grows drastically near $p_f=1-1/N$. Inset: Plot of the relationship between $t$ and $p$. $t$ increases drastically but in a power-law manner as $p$ approaches $p_f$. Simulations are carried out with $N=8,000$ mono particles at $t=0$ on $400\times 400$ square lattices.}
\label{fig1}
\end{figure}

When we study a PT problem of networks, control parameter is the number of edges added to the system per the total number of nodes. Following this convention, we introduce a new variable $p$, which is defined as the number of cluster aggregation events per the total particle number. Whenever two clusters merge, $p$ is increased by $\delta p=1/N$. Since $N-1$ aggregation events occur during all aggregation processes, the aggregation event stops at $p_f=1-1/N$. $p$ depends on $t$ in a nonlinear way as shown in the inset of Fig.~\ref{fig1}(b).

In the original study of the DLCA model, the PT was not studied because the giant cluster size $G(t)$ increases monotonically as the time $t$ increases (Fig.~\ref{fig1}(a)). However, we show here that when the giant cluster size $G$ is traced as a function of $p$, it drastically increases, exhibiting a discontinuous PT, as shown in Fig.~\ref{fig1}(b). This different behavior is caused by the nonlinear relationship between $t$ and $p$ shown in the inset of Fig.~\ref{fig1}(b). When $p$ is small, $t$ increases almost linearly with respect to $p$. However, as $p$ approaches to $p_f$, $t$ increases drastically in a power-law manner. As a result, $G$ exhibits a discontinuous PT.

\begin{figure}[t]
\includegraphics[width=1.0\linewidth]{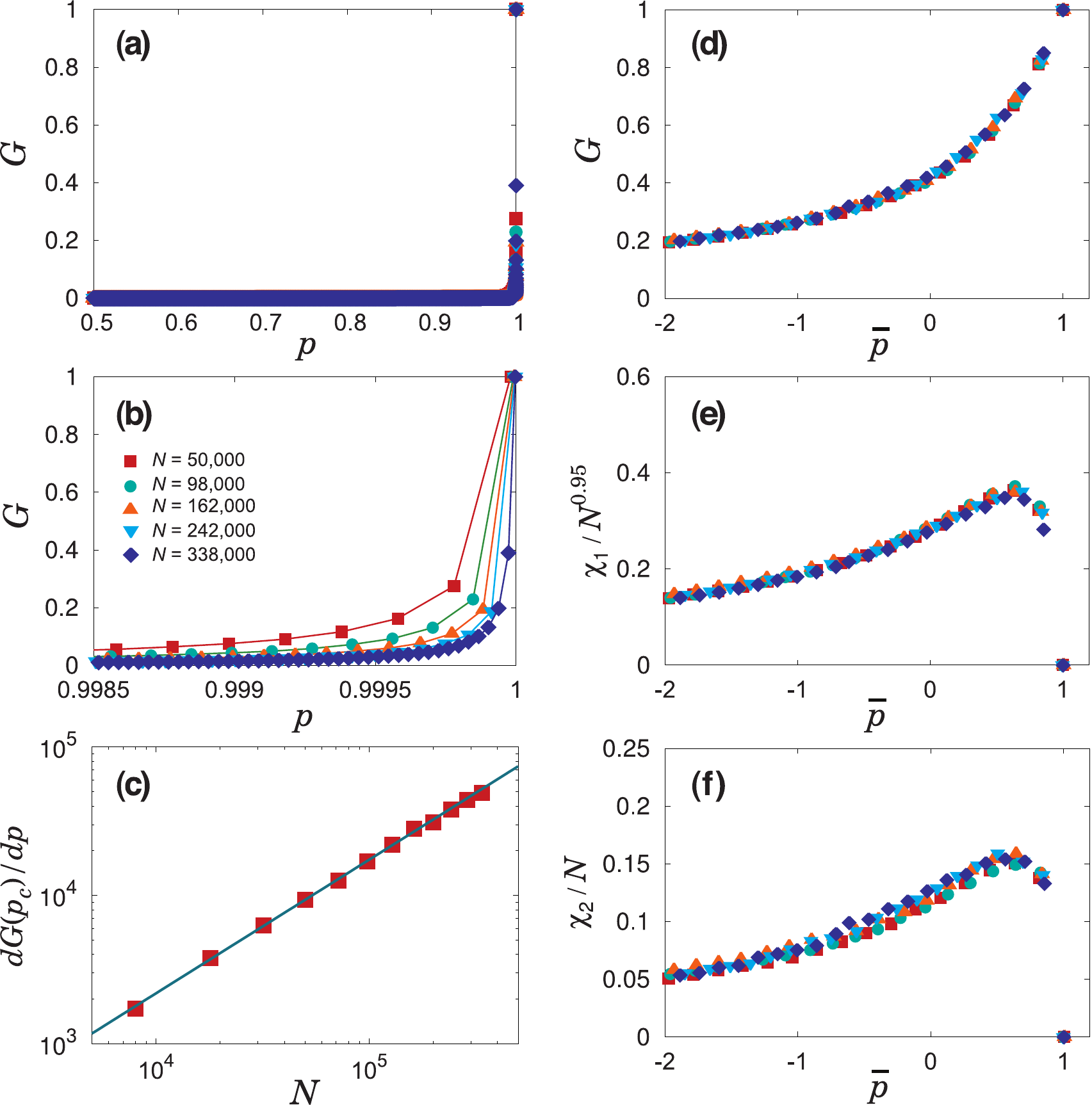}
\caption{(color online) (a) Plot of $G_N$ versus $p$ for different particle numbers, $N=50,000$ ($\square$), $N=98,000$ ($\circ$), $N=162,000$ ($\triangle$),
$N=242,000$ ($\bigtriangledown$), and $N=338,0000$ ($\diamond$) at a fixed density $\rho=0.05$. $G_N$ begins to increase drastically near $p\approx 1$ and thus the data are not distinguishable for different sizes in the region far smaller than $p=1$.
(b) Zoom-in plot of $G_N$ versus $p$ near $p_c$. As the system size grows, the giant cluster grows more drastically. (c) Plot of $dG_N(p)/dp$ calculated at $p_c(N)$ versus $N$. The slope $dG_N(p_c)/dp$ increases as $N^{0.86 \pm 0.02}$, indicating that it diverges in the thermodynamic limit.
(d) Plot of $G$ versus $\overline{p} \equiv (p-p_{d})dG_N(p_{c})/dp$ for different $N$. $p_d(N)$ is the $p$-intercept of the tangent of the curve $G_N(p)$ at $p_c$. (e) Plot of $\chi_{1}/N^{0.95}$ versus $\overline{p}$. (f) Plot of $\chi_{2}/N$ versus $\overline{p}$.
For (d), (e), and (f), the data collapse well onto a single curve.}
\label{fig2}
\end{figure}

To check if the PT is indeed discontinuous, we use the finite-size scaling theory recently proposed for studying the explosive PT~\cite{fss}. For the first, we measure $G_N(p)$ for different $N$ under the condition that the particle density $\rho$ remains fixed. The results are shown in Fig.~\ref{fig2}(a) and (b). For a given $N$, we pick up a $p_c(N)$ at which the increasing rate $dG_N/dp$ is maximum. Then, the $p$-intercept of the tangent of $G_N(p)$ at $p_c$, denoted by $p_d$, is determined as
\begin{equation}
p_d=p_c-\Big(\frac{dG_N(p)}{dp}\Big|_{p_c}\Big)^{-1}G_N (p_c).
\end{equation}
Then, $p_d$ also depends on $N$. We find that $dG_N(p)/dp$ at $p_c$ increases in a power-law manner as $\sim N^{0.86}$ (Fig.~\ref{fig2}(c)), indicating that the giant cluster size increases more drastically as $N$ increases. {\it Thus, the transition is indeed discontinuous.} Near $p_d(N)$, $p$ is rescaled as $\overline{p}=(p-p_{d})dG_N(p_{c})/dp$, which is then $N$-independent. The giant cluster is then plotted as a function of $\overline{p}$. Indeed, we can see that the curves of the giant cluster sizes for different $N$ collapse well onto a single curve (Fig.~\ref{fig2}(d)). Thus, the order parameter of the PT is written in the scaling form,
\begin{equation}
G(p)\propto N^{-\beta/{\bar \nu}}f_0((p-p_d)N^{1/{\bar \nu}}),
\end{equation}
where $f_0(x)$ is a scaling function, and $\beta=0$ and $1/{\bar \nu}\approx 0.86\pm 0.02$. We remark that this finite-size scaling form differs from the conventional one used in the continuous transition in the aspect that $p_d$ depends on the particle number $N$, which does in turn the system size $L$. Whereas, in the conventional scaling form used for a continuous PT, $p_d$ is replaced by $p_c(\infty)$, i.e., the critical point in the thermodynamic limit, which is independent of $N$. For comparison, the modified ER models~\cite{mendes, hklee, science2} which were claimed to exhibit continous PTs show $\beta/{\bar \nu} > 0$ even though their values are extremely small.

We examine the behavior of the susceptibility. The susceptibility is defined in two ways. The first is the mean cluster size $\chi_1(p)\equiv \sum^{\prime}_s s^2 n_s(p)/\sum_s^{\prime}sn_s(p)$, which exhibits a peak at $p_{c1}$. This quantity approaches $p_f$ defined earlier, as $N$ increases. The susceptibility at $p_{c1}$ increases with $N$ as $\chi_1(p_{c1}(N))\sim N^{0.95 \pm 0.01}$. Thus, $\chi_1(p)$ is written in the scaling
form $\chi_1(p)\sim N^{\gamma_1/{\bar \nu}}f_1((p-p_d)N^{1/{\bar \nu}})$, where $f_1(x)$ is another scaling function and $\gamma_1/{\bar \nu}\approx 0.95 \pm 0.01$. The scaling behavior is confirmed numerically in Fig.~\ref{fig2}(e). The other susceptibility is the fluctuation of the giant component
sizes, i.e., $\chi_2(p)\equiv N\sqrt{\langle G_N^2(p)\rangle-\langle G_N(p)\rangle^2}$. This quantity exhibits a peak at $p_{c2}$. We find that $\chi_2(p_{c2}(N))\sim N$. Thus, $\gamma_2/{\bar \nu} = 1$ and $\chi_2(p)\sim Nf_2((p-p_d)N^{1/{\bar \nu}})$ with a scaling function $f_2$. The scaling behaviors are also confirmed numerically in Fig.~\ref{fig2}(f).

The cluster aggregation process may be described via an asymmetric Smoluchowski equation:
\begin{equation}
\frac{dn_s}{dp}=\sum_{i+j=s}\frac{k_ik^{\prime}_j}
{C(p)C^{\prime}(p)}n_i n_j-\frac{n_sk_s}{C(p)}-\frac{n_sk^{\prime}_s}{C^{\prime}(p)},
\end{equation}
where $n_s \equiv N_s/N$ is the concentration of $s$-sized clusters, which depends on $p$, and $k_ik^{\prime}_j/(CC^{\prime})$ is a collision kernel, where $C(p) \equiv \sum_{i} k_{i}n_{i}$ and $C^{\prime}(p) \equiv \sum_{i} k^{\prime}_{i}n_{i}$. $k_{i}/C$ and $k^{\prime}_{j}/C^{\prime}$ are the probabilities for $i$- and $j$-sized clusters to merge, in which the prime denotes mobile clusters and the other does immobile clusters in simulations. Their kernels are different below. 

We measure size-dependent behaviors of $k_{s}/C$ and $k^{\prime}_{s}/C^{\prime}$ numerically. Even though it is not manifest that $k_{s}/C$ and $k^{\prime}_{s}/C^{\prime}$ follows a power law for the Brownian particle case (see Fig.~\ref{fig4}(a)), we roughly estimate that $k_{i}/C \sim i^{0.3}$ and $k^{\prime}_{j}/C^{\prime}\sim j^{-0.2}$. The relation between these two exponents is mentioned later.
With these exponent values, we solve the asymmetric Smoluchowski equation numerically and obtain that $\beta=0$, $1/{\bar \nu} = 1$, 
$\gamma_{1}/{\bar \nu} = 1$ and $\gamma_{2}/{\bar \nu} = 1$. These obtained numerical values indicate more clearly that the transition is indeed discontinuous.  

\begin{figure}[t]
\includegraphics[width=1.0\linewidth]{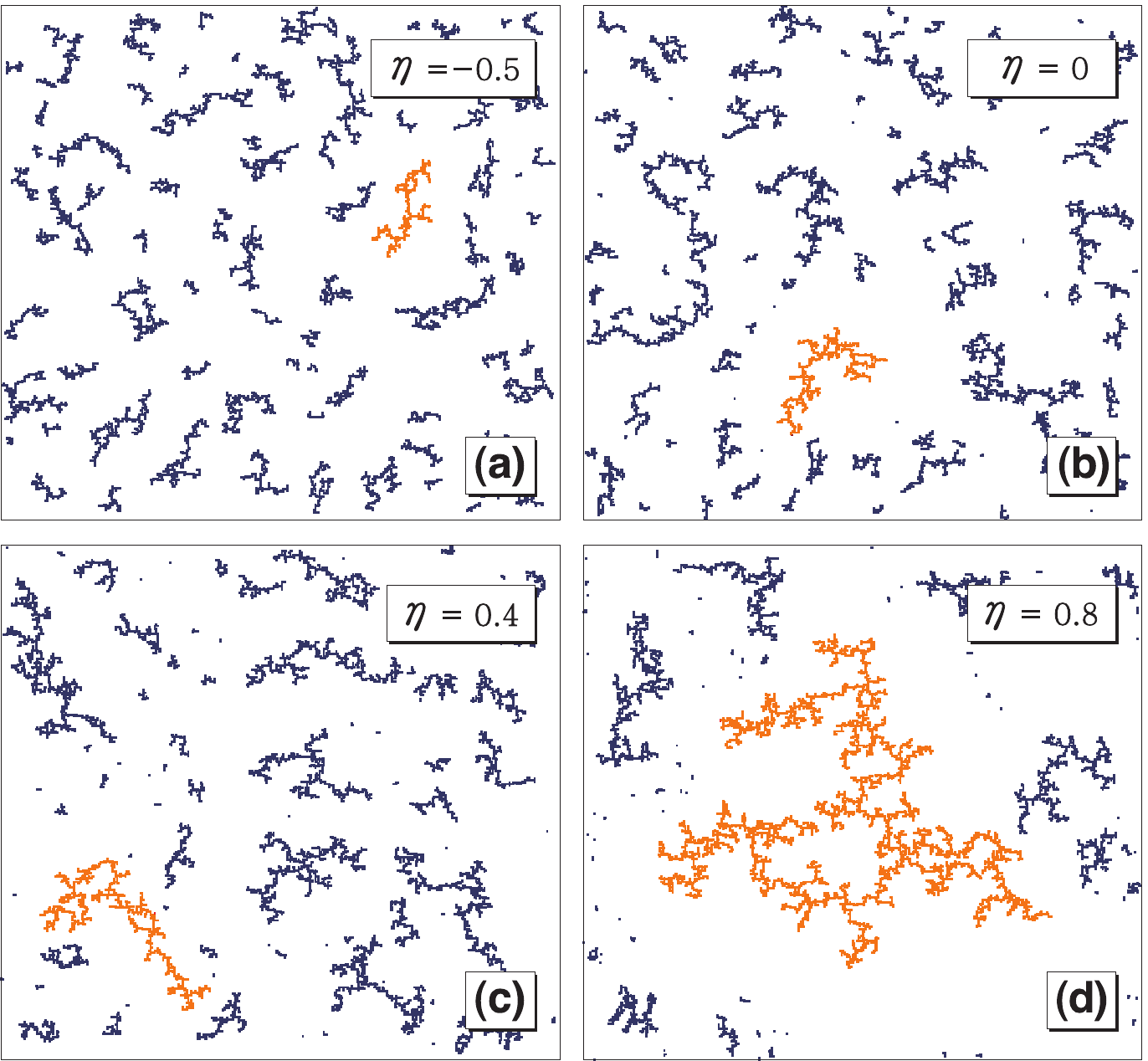}
\caption{(color online) Snapshots of the system for various values of $\eta$ at $p=0.99$. The velocity of each cluster is given as $v_{s} \propto s^{\eta}$, where $s$ is cluster size. Numerical simulations are carried out for $N = 8,000$ particles on $L \times L = 400 \times 400$ square lattices. Since $p$ is fixed, the number of clusters for each case is equal. The giant cluster is represented in a different color (gray, orange). The cluster-size distribution becomes more heterogeneous as $\eta$ increases.}
\label{fig3}
\end{figure}

We now consider a more general case in which the velocity is given as $v_s \propto s^{\eta}$~\cite{cdist}, where $s$ is cluster size. To implement this case, a cluster is picked up with a probability proportional to $s^{\eta}$. The other rules in the numerical simulations remain the same. When clusters merge, the variable $p$ is advanced by $1/N$, regardless of the cluster size $s$. The time is by $\delta t=1/(\sum_s N_s s^{\eta})$.
Intuitively, when $\eta$ is small or negative, fewer large-sized clusters are selected, so their growth is suppressed. Medium-sized clusters are
abundant even close to $p_f$, and they merge suddenly. In this case, a discontinuous PT takes place. In contrast, when $\eta$ is positively large, 
more large-sized clusters are selected and they have more chance of colliding with other clusters, and merge to a bigger cluster. Thus, they can grow faster than smaller clusters can, so the giant cluster grows continuously and the PT is continuous. Thus we expect that there is a tricritical point $\eta_{c}$ across which the transition type is changed. We show snapshots of the system for different values of $\eta$ at $p=0.99$ in Fig.~\ref{fig3}, which support the above argument.

\begin{figure}[t]
\includegraphics[width=1.0\linewidth]{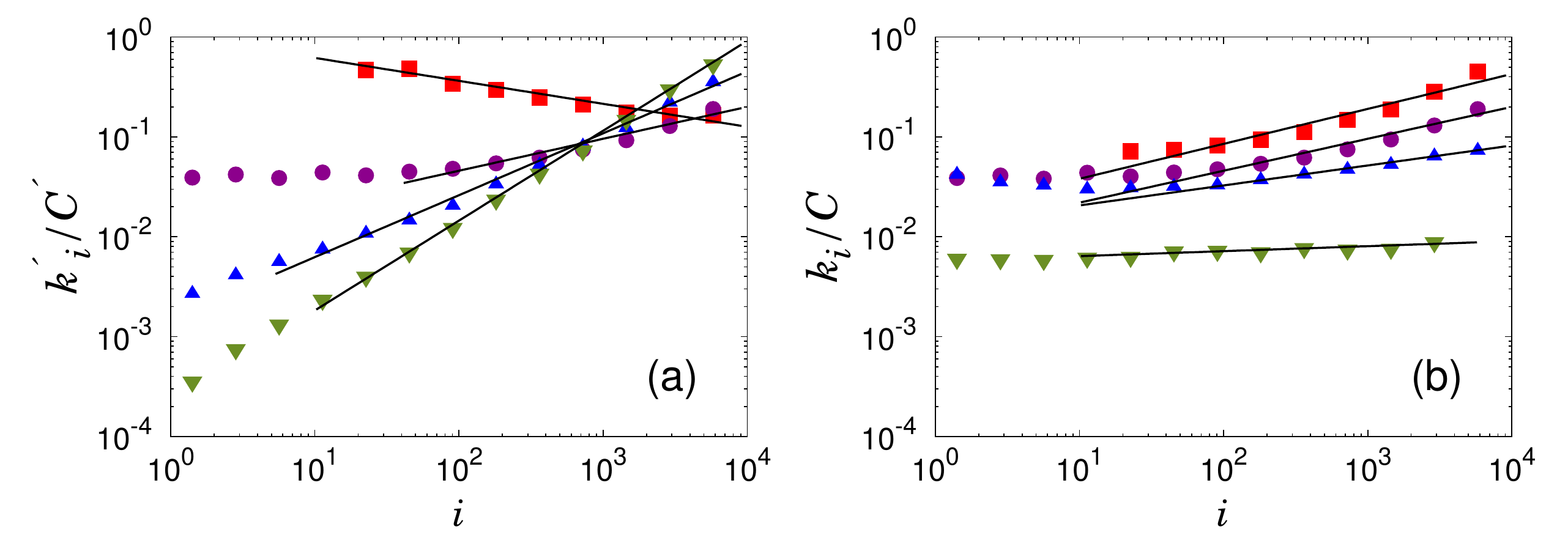}
\caption{(color online) (a) Numerical estimations 
of $k^{\prime}_{i}/C^{\prime}$ at $p_{d}$ for $\eta=-0.5$ $(\square)$, $\eta=0$ $(\circ)$, 
$\eta=0.4$ $(\triangle)$ and $\eta=0.8$ ($\triangledown$). Simulations are
carried out for $N=8,000$ and $L=400$.
Slopes of the guidelines are $-0.23 \pm 0.02$ for ($\eta=-0.5$), $0.32 \pm 0.04$ ($\eta=0$), $0.62 \pm 0.01$ ($\eta=0.4$), and $0.88 \pm 0.01$ ($\eta=0.8$).
(b) Numerical estimations of the $k_{i}/C$ at $p_{d}$. The same symbols are used as (a). System size is $N=8000$ and $L=400$. Slopes of the guidelines are $0.35 \pm 0.04 $ ($\eta=-0.5$), $0.32 \pm 0.04$ ($\eta=0$), $0.2 \pm 0.01$ ($\eta=0.4$),
and $0.05 \pm 0.005$ ($\eta=0.8$).} 
\label{fig4}
\end{figure}

It has been roughly argued that the collision kernel in the Smoluchowski equation may be related to the perimeter of a cluster as $k_{i} \sim i^{1-1/d_f}$~\cite{ernst}. However, when the selection probability is taken into consideration, the collision kernel of mobile clusters 
may be modified as $k^{\prime}_{i} \sim i^{\eta+1-1/d_f}$.
For the Brownian case with $\eta=-0.5$, by using the fractal dimension $d_f=1.4-1.5$, the measured values $k^{\prime}_i\sim i^{-0.2}$ for mobile clusters and $k_i\sim i^{0.3}$ for immobile clusters (Fig.~\ref{fig4}(a) and (b)) are reasonable.  
When $\eta$ is sufficiently large such as $\eta > 0.8$, the largest cluster grow by merging small-sized clusters. In such a merging process, small-sized clusters can penetrate into the interior of the giant cluster, and then the merging probability can be independent of cluster size (Fig.~\ref{fig4}(b)). Thus, $k^{\prime}_i\sim i^{\eta}$ and $k_i\sim$ constant. 

\begin{figure}[t]
\includegraphics[width=1.0\linewidth]{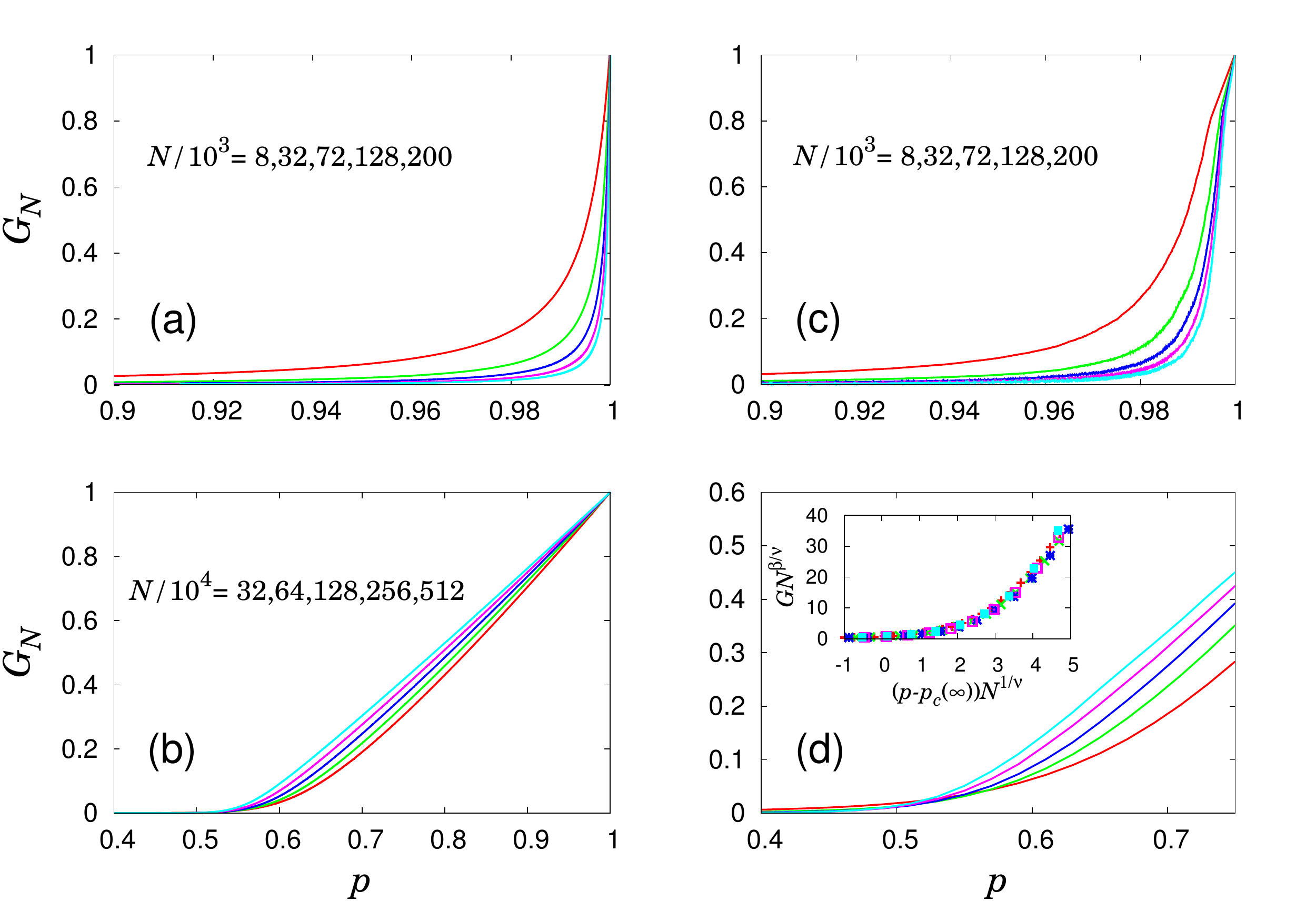}
\caption{(color online) Plot of $G_{N}$ versus $p$ for the Smoluchowski equation with
$k_i \sim 1$ and $k^{\prime}_j \sim j^{0.8}$ (a), and
$k_i \sim 1$ and $k^{\prime}_j \sim j^{1.5}$ (b). 
As the system size grows, $G_N$ grows more drastically and $p_c(N)$ approaches one in (a), but it decreases 
to $p_c(\infty)>0$ in (b). For (b), the transition 
is continuous. Similar plot for the DLCA 
with $\eta = 0.8$ (c) and $\eta = 1.5$ (d). 
Similar behaviors are shown. Simulations are 
performed for $N/10^{3} = 8, 32, 72, 128$ and $200$.
Inset of (d): Finite size scaling of $G_N(p)$ for different system sizes with $1/\nu=0.28$, 
$\beta/\nu=0.5$ and $p_c({\infty})=0.42$.
These numerical values depend on $\eta$. Thus, for $\eta > \eta_c$, the critical points for different $\eta$ form a critical line.}
\label{fig5}
\end{figure}

We integrate the Smoluchowski equation with using $k^{\prime}_{i}/C^{\prime} \sim i^{\eta}$ and $k_{i}/C \sim 1$. 
We find that the transition behavior changes across $\eta_c \approx 1.3-1.4$. When $\eta < \eta_c$ 
($\eta > \eta_c$), as the system size increases, the percolation threshold increases to one (decreases to a finite percolation threshold). Those behaviors can be observed in Fig.~\ref{fig5}(a) and (b). Moreover, for the latter case, the transition turns out to be continuous. 
Therefore we conclude that there exists a tricritical point which locates at $\eta_c$. Similar behavior is observed for the DLCA model (Fig.~\ref{fig5}(c) and (d)). We also check the cluster size distribution at the percolation threshold. Indeed, the distribution obeys a power law 
$n_s \sim s^{-\tau}$. Also, the exponent 
satisfies $\tau<(>)2$ for $\eta<(>) \eta_c \approx 1.3-1.4$. The exponent $\tau=2$ is marginal between the continuous and the discontinuous PT, 
which was proven analytically in the cluster aggregation model~\cite{can}. Therefore our numerical result confirm that the tricritical point 
locates near $\eta_c\approx 1.3-1.4$.  

In summary, we have studied the DLCA model as an example of real-world systems exhibiting discontinuous PT. The velocity of an $s$-sized cluster is given in a general form $v_s \sim s^{\eta}$. When $\eta < (>) \eta_c$, the PT is discontinuous (continuous), where a tricritical point $\eta_c$ is roughly estimated to be $1.3-1.4$. Since the case $\eta=-0.5$ corresponds to the Brownian particle motion in real-world systems, we can say that a discontinuous PT can take place in real-world non-equilibrium systems. We also introduced and studied an asymmetric Smoluchowski equation, and determined the tricritical point from the fact that the cluster size distribution follows a power-law behavior with exponent $-2$ at the tricritical point. We finally remark that the discovery of the explosive PT in the DLCA model was made by tracing the giant cluster size as a function of $p$ instead of $t$, indicating that a discontinuous PT may be explored as a function of a unconventional parameter.

We thank D. Kim for helpful discussion. This study was supported by the NRF grants funded by the MEST (Grant No. 2010-0015066) and the NAP of KRCF (BK), and the Seoul Science Foundation and the Global Frontier program (YSC).

\end{document}